\documentclass[twocolumn,showpacs,superscriptaddress,prl,nofootinbib]{revtex4-1}

\usepackage[latin9]{inputenc}
\usepackage{amsmath}
\usepackage{amssymb}
\usepackage{amsthm}
\usepackage{graphicx}
\usepackage{color}

\theoremstyle{plain}
\newtheorem*{theorem*}{Theorem}

\makeatletter

\@ifundefined{textcolor}{}
{
 \definecolor{BLACK}{gray}{0}
 \definecolor{WHITE}{gray}{1}
 \definecolor{RED}{rgb}{1,0,0}
 \definecolor{GREEN}{rgb}{0,1,0}
 \definecolor{BLUE}{rgb}{0,0,1}
 \definecolor{CYAN}{cmyk}{1,0,0,0}
 \definecolor{MAGENTA}{cmyk}{0,1,0,0}
 \definecolor{YELLOW}{cmyk}{0,0,1,0}
}

\usepackage{color}
\usepackage{amsfonts}
\usepackage{graphics}
\usepackage{bm} 
\usepackage{epsfig}\usepackage{amsthm}

\usepackage{lipsum}

\newcommand\blfootnote[1]{%
  \begingroup
  \renewcommand\thefootnote{}\footnote{#1}%
  \addtocounter{footnote}{-1}%
  \endgroup
}

\def\identity{\leavevmode\hbox{\small1\kern-3.8pt\normalsize1}}

\renewcommand{\epsilon}{\varepsilon}

\makeatother

\begin{document}

\title{Observable quantum entanglement due to gravity}

\author{Tanjung Krisnanda}
\affiliation{School of Physical and Mathematical Sciences, Nanyang Technological University, 637371 Singapore, Singapore}
\blfootnote{Correspondence: T. Krisnanda (tanjungkrisnanda@gmail.com) or T. Paterek (tomasz@paterek.info)}

\author{Guo Yao Tham}
\affiliation{School of Physical and Mathematical Sciences, Nanyang Technological University, 637371 Singapore, Singapore}

\author{Mauro Paternostro}
\affiliation{School of Mathematics and Physics, Queen's University, Belfast BT7 1NN, United Kingdom}

\author{Tomasz Paterek}
\affiliation{School of Physical and Mathematical Sciences, Nanyang Technological University, 637371 Singapore, Singapore}
\affiliation{MajuLab, International Joint Research Unit UMI 3654, CNRS, Universite Cote d'Azur, Sorbonne Universite, National University of Singapore, Nanyang Technological University, Singapore}

\begin{abstract}
No experiment to date has provided evidence for quantum features of the gravitational interaction.
Recently proposed tests suggest looking for the generation of quantum entanglement between massive objects as a possible route towards the observation of such features.
Motivated by advances in optical cooling of mirrors, here we provide a systematic study of entanglement between two masses that are coupled gravitationally.
We first consider the masses trapped at all times in harmonic potentials (optomechanics) and then the masses released from the traps.
This leads to the estimate of the experimental parameters required for the observation of gravitationally induced entanglement. 
The optomechanical setup demands LIGO-like mirrors and squeezing or long coherence times,
but the released masses can be light and accumulate detectable entanglement in a timescale shorter than their coherence times.
No macroscopic quantum superposition develops during the evolution.
We discuss the implications from such thought experiments regarding the nature of the gravitational coupling.
\end{abstract}

\maketitle

\noindent{\bf INTRODUCTION}

\noindent The successful unification of electromagnetic, weak and strong interactions within the quantum framework strongly suggests that gravity should also be quantised.
Up to date, however, there is no experimental evidence of quantum features of gravity.
In numerous experiments gravity is key to the interpretation of the observed data, 
but it is sufficient to use Newtonian theory (quantum particle moving in a background classical field) or general relativity (quantum particle moving in a fixed spacetime) to gather a meaningful understanding of such data.
Milestone experiments described within Newtonian framework include gravity-induced quantum phase shift in a vertical neutron interferometer~\cite{gphase}, 
precise measurement of gravitational acceleration by dropping atoms~\cite{gravimeter}, 
or quantum bound states of neutrons in a confining potential created by the gravitational field and a horizontal mirror \cite{nfall}.
Quantum experiments that require general relativity include gravitational redshift of electromagnetic radiation~\cite{gshift1} or time dilation of atomic clocks at different heights~\cite{clock1}.

A number of theoretical proposals discussed scenarios capable of revealing quantumness of gravity.
For example Refs.~\cite{massg1,massg2,massg22,massg3,massg4,massg5,massg6,cavenexp,qgdis} proposed the observation of a probe mass interacting with the gravitational field generated by another mass.
More recent proposals put gravity in a role of mediator of quantum correlations and are based on the fact that quantum entanglement between otherwise non-interacting objects can only increase via a quantum mediator \cite{revealing,gravity1,gravity2}.
Motivated by these proposals and by advances in optomechanics \cite{optoreview}, in particular the cooling of massive mechanical (macroscopic) oscillators close to their quantum ground state \cite{ligomirror,nm,nanomirror} 
and the measurement of quantum entanglement of a two-mode system~\cite{evalue1,evalue2,evalue3}, we study two nearby cooled masses interacting gravitationally.

We propose two scenarios capable of increasing gravitational entanglement between masses.
In the first scenario, we consider masses trapped at all times in 1D harmonic potentials (optomechanics). In the second one, the masses are released from the optical traps.
For both settings, we derive an analytic figure of merit characterising the amount of gravitationally induced entanglement and the time it takes to observe it.
The derivation includes various initial states and shows that the objects have to be cooled down very close to their ground states and that squeezing of their initial state significantly enhanced the amount of generated entanglement.
We then formulate a numerical approach, which accounts for all the relevant sources of noise affecting the settings that we propose, to identify a set of parameters required for the observation of such entanglement. 
Finally, we discuss the conclusions that can be drawn from this experiment with emphasis on the need for independent laboratory verification that the gravitational interaction between nearby objects is indeed mediated.

\vspace{0.5cm}
\noindent{\bf RESULTS}

\noindent{\bf A. Proposed setup}

\begin{figure}[t]
\centering
\includegraphics[width=0.48\textwidth]{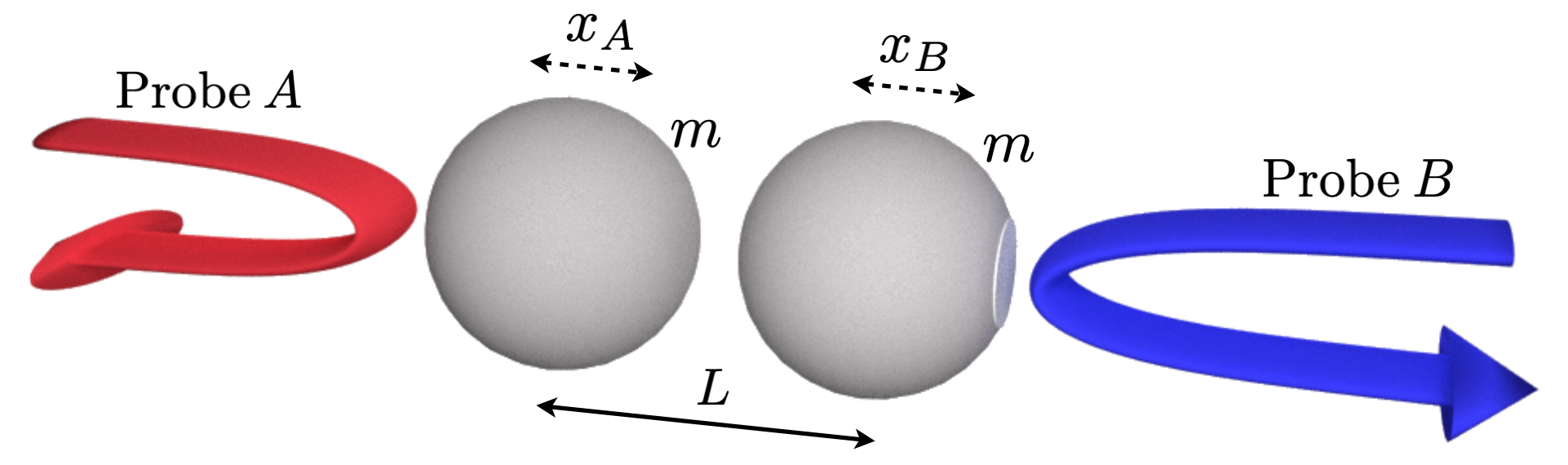}
\caption{Proposed experimental setup.
Two masses, placed at a distance $L$, are either trapped with harmonic potentials at all times or released after cooling has been achieved.
The particles are assumed to be cooled down near the ground state of their trapping potentials.  
We study entanglement generated in both scenarios and note that it can be probed with weak light fields.
Our model includes gravitational coupling (dominant), noise, damping, decoherence and Casimir forces.}
\label{FIG_setup}
\end{figure}

\noindent Consider two particles, separated by a distance $L$, as depicted in Fig. \ref{FIG_setup}. 
In what follows, we study the setting where the massive particles are either held or released from unidimensional harmonic traps.
In the former case one can treat the particles as identical harmonic oscillators, with the same shape, mass $m$, and vibrational frequency $\omega$. 
The two oscillators and the gravitational interaction between them give rise to the total Hamiltonian $H=H_0+H_{\text g}$, where
\begin{equation}\label{EQ_hamiltonian}
H_0=\frac{p_A^2}{2m}+\frac{1}{2}m\omega^2 x_A^2+\frac{p_B^2}{2m}+\frac{1}{2}m\omega^2 x_B^2
\end{equation}
and $H_{\text g}$ describes the gravitational term.
If the harmonic traps are removed the corresponding Hamiltonian simplifies to $H_0=(p_A^2+p_B^2)/2m$.
Before we proceed with detailed calculations, we shall discuss generic features of the gravitational term and the conditions required for the creation of entanglement.

In general, the gravitational term $H_{\text g}$ depends on the geometry of the objects. 
Various configurations have been analysed in the Supplementary Information accompanying this paper. 
The results of such analysis suggest that spherical masses give rise to the highest amount of generated entanglement.
The Newtonian gravitational energy of this setting is the same as if the two objects were point-like masses, that is 
$H_{\text g}=-Gm^2/ (L + x_B - x_A)$, where $L$ is the distance between the objects at equilibrium and $x_A$ ($x_B$) is the displacement of mass $A$ ($B$) from equilibrium.
By expanding the energy in the limit $x_A-x_B \ll L$, which is well justified for oscillators that are cooled down close to their ground state, one gets
\begin{equation}\label{EQ_hspheresexp}
H_{\text g}=-\frac{Gm^2}{L}\left(1+\frac{(x_A - x_B)}{L}+\frac{(x_A - x_B)^2}{L^2}+\cdots\right).
\end{equation}
The first term is a rigid energy offset, while the second is a bi-local term and cannot thus give rise to quantum entanglement. 
The third term, which is proportional to $(x_A-x_B)^2$, is the first that couples the masses. 
When written in second quantisation, it becomes apparent that this term includes contributions responsible for the correlated creation of excitations in both oscillators. 
In the quantum optics language, this is commonly referred to as a ``two-mode squeezing" operation, which can in principle  entangle the masses provided a sufficient strength of their mutual coupling.
Based on this observation we provide an intuitive argument setting the scales of experimentally relevant parameters, which will then be proven rigorously.

\vspace{0.5cm}
\noindent{\bf B. Calculations of entanglement: Oscillators}

\noindent In order to achieve considerable entanglement, we should ensure that the coupling (third term) in Eq.~(\ref{EQ_hspheresexp}) is comparable to the energy $\hbar \omega$ of each oscillator, that is $Gm^2(x_A - x_B)^2/L^3 \sim \hbar \omega$.
As we assume that the oscillators are near their ground state, we estimate their displacements by the ground state extension, $(x_A-x_B)^2 \sim 2 \hbar / m \omega$. 
We thus introduce the (dimensionless) figure of merit
\begin{equation}
\eta \equiv \frac{2 G m}{\omega^2L^3}.
\label{EQ_OSC_MERIT}
\end{equation}
We should have $\eta\sim1$ in order for the oscillators to be significantly entangled. 
This sets the requested values of the experimentally relevant parameters $m$, $\omega$, and $L$.

In what follows we will demonstrate the following results, which embody the key findings of our investigation:
(i) Starting from the ground state of each oscillator and assuming (for the sake of argument) only negligible environmental noise, the maximum entanglement (as quantified by the logarithmic negativity~\cite{negativity,adesso2004extremal}) generated during the dynamics is given by $E_{\mbox{\scriptsize th}}^{\max}\approx \eta/\ln{2}$. 
Moreover, the time taken for entanglement to reach such maximum value is $t_{\mbox{\scriptsize th}}^{\max}=\pi/2(1-\eta)\omega$;
(ii) Single-mode squeezing of the initial ground state of each oscillator substantially enhances the gravity-induced entanglement. 
The corresponding maximum entanglement becomes $E_{\mbox{\scriptsize sq}}^{\max}\approx |s_A+s_B|/\ln{2}$, 
where $s_j~(j=A,B)$ is the degree of squeezing of the $j^\text{th}$ oscillator, and we assume $\eta \ll s_A, s_B$.
In this case, the maximum entanglement is reached in a time $t_{\mbox{\scriptsize sq}}^{\max}=\pi/2\eta\omega$;
(iii) Weaker entanglement is generated with increasing temperature of the masses or coupling to the environment.

As the third term in Eq.~(\ref{EQ_hspheresexp}) is already very small under usual experimental conditions,\footnote{Note that the ratio between any two consecutive terms in Eq.~(\ref{EQ_hspheresexp}) is given by $(x_A - x_B) / L \sim \sqrt{\hbar / m \omega L^2}$. 
For instance, taking $m=100$~$\mu$g, $\omega=100$~kHz, and $L=0.1$~mm gives this ratio $\sim 10^{-12}$, and for macroscopic values $m = 1$~kg, $\omega = 0.1$~Hz, and $L = 1$~cm the ratio is $\sim10^{-15}$.}
we neglect all terms of order higher than the second in the displacement from equilibrium. 
We note Ref. \cite{linearp} for similar treatment of linearised central-potential interactions.
By taking the total Hamiltonian with a suitably truncated gravitational term $H_{\text g}$, one gets a set of Langevin equations in Heisenberg picture
\begin{equation}\label{EQ_langevins}
\begin{aligned}
\dot { X}_j &=\omega \:  P_j\qquad (j=A,B),\\ 
\dot { P}_A&=-\omega\left(1-\eta\right) X_A-\omega\eta\, X_B-\gamma \,  P_A+\xi_A+ \nu,\\
\dot { P}_B&=-\omega\left(1-\eta\right) X_B-\omega\eta\, X_A-\gamma \,  P_B+\xi_B-\nu,\\
\end{aligned}
\end{equation}
where we have introduced the constant frequency $\nu={Gm^2}/\sqrt{\hbar m\omega L^4}$ and the dimensionless quadratures $X_j=\sqrt{m\omega/\hbar}\:x_j$ and $P_j=p_j/\sqrt{\hbar m\omega}$. 
These equations incorporate Brownian-like noise -- described by the noise operators $\xi_j$ -- and damping (at rate $\gamma$) affecting the dynamics of the mechanical oscillators, due to their interactions with their respective environment.
We assume the (high mechanical quality) conditions $\mathcal{Q}=\omega/\gamma \gg 1$, as it is the case experimentally, so that the Brownian noise operators can {\it de facto} be treated as uncolored noise and we can write $\langle \xi_j(t)\xi_j(t^{\prime})+\xi_j(t^{\prime})\xi_j(t)\rangle/2 \simeq \gamma(2 \bar n+1)\delta(t-t^{\prime})$ for $j=A,B$~\cite{brownian1,brownian2}. Here, $\bar n=(e^\beta-1)^{-1}$ is the thermal phonon number with $\beta=\hbar \omega/k_B T$ and $T$ the temperature of the environment with which the oscillators are in contact. 

The linearity of Eqs.~(\ref{EQ_langevins}) and the Gaussian nature of the noise make the theory of continuous variable Gaussian systems very well suited to the description of the dynamics and properties of the oscillators under scrutiny. In this respect, the key tool to use is embodied by the covariance matrix $V(t)$ associated with the state of the system, whose elements $V_{ij}(t) = \langle u_i(t)u_j(t)+u_j(t)u_i(t)\rangle/2-\langle u_i(t)\rangle \langle u_j(t)\rangle$ encompass the variances and correlations of the elements of the quadrature vector $u(t)=(X_A(t), P_A(t), X_B(t), P_B(t))^T$.
The temporal behaviour of physically relevant quantities for our system of mechanical oscillators can be drawn from $V(t)$ by making use of the approach for the solution of the dynamics that is illustrated in Methods.

Due to weakness of the gravitational coupling, we have $\eta \ll 1$ in practically any realistic experimental situation, and we thus assume such conditions throughout.
In the case of no damping (i.e., $\gamma = 0$) and assuming an initial (uncorrelated) thermal state of the oscillators, 
a tedious but otherwise straightforward analytical derivation shows that the entanglement between the mechanical systems, as quantified by the logarithmic negativity, oscillates in time with an amplitude of $\eta/\ln{2}-\log_2(2\bar n+1)$. At low operating temperature, a condition achieved through a combination of passive and radiation-pressure cooling~\cite{optoreview}, $\bar n\approx 0$ and the maximum entanglement between the oscillators is $E_{\mbox{\scriptsize th}}^{\max}\approx \eta/\ln{2}$, a value reached at a time $t^{\max}_{\mbox{\scriptsize th}}$ such that $\omega t^{\max}_{\mbox{\scriptsize th}}=\pi/2(1-\eta)$. 

An analytic solution is also possible for the case of mechanical systems initially prepared in squeezed thermal states, a situation that can be arranged by suitable optical driving~\cite{Vanner2013,Rashid}.
Each mass is prepared in a state $S\rho_{\text {th}}S^{\dagger}$, where $\rho_{\text {th}}$ is a thermal state and $S=\exp{(-i\:s(X^2-P^2)/2)}$ is the squeezing operator with strength $s$. 
This operator corresponds to anti-squeezing (squeezing) the position quadrature for $s>0$ ($s<0$).
By writing individual-oscillator squeezing as $s_j$ and assuming $s_j \gg \eta$, the entanglement is again observed to oscillate, but with amplitude $|s_A + s_B|/\ln{2}-\log_2(2\bar n+1)$. 
Note that it is irrelevant whether the quadratures of both masses are squeezed or anti-squeezed.
We provide an explanation in the Supplementary Information.
Therefore, only the degree of pre-available single-oscillator squeezing and the environmental temperature set a limit to the amount of entanglement that can be generated between the mechanical systems through the gravitational interaction.  
In the low temperature limit, where $E_{\mbox{\scriptsize sq}}^{\max}\approx |s_A+s_B|/\ln{2}$, which is in principle arbitrarily larger than the case without squeezing, 
a time $t^{\max}_{\mbox{\scriptsize sq}}=\pi/(2\eta\omega)\gg t^{\max}_{\mbox{\scriptsize th}}$ would be required for such entanglement to accumulate. 
Needless to say, long accumulation times are far from the  possibilities offered by state-of-the-art optomechamical experiments, which prompts an assessment that includes ab initio the effects of environmental interactions.

In the case of noisy dynamics, however, an analytical solution is no longer available and we have to resort to a numerical analysis.
Let us therefore consider the figure of merit $\eta$ in order to set parameters for numerical investigation.
We consider two oscillators of spherical shape with uniform density $\rho$ and radius $R$, which are separated by a distance $L = 2.1 R$. This might be a situation matching current experiments in levitated optomechanics~\cite{kiesel,Vovrosh}, which are rapidly evolving towards the possibility of trapping multiple dielectric nano-spheres in common optical traps and controlling their relative positions. 
However, low-frequency oscillators, which are favourable for the figure of merit and typically associated with large masses, are unsuited to such platforms and would require a different arrangement, such as LIGO-like ones~\cite{ligomirror}.

In terms of the density $\rho$, we have $\eta=8\pi G\rho/3(2.1)^3 \omega^2$, which does not depend on the dimensions of the oscillators nor their mass.
As the density of materials currently available for such experiments varies within a range of only two orders of magnitude, the linear dependence on $\rho$ sets a considerable restriction on the values that $\eta$ can take. 
The densest naturally available material is Osmium, which has $\rho=22.59$~g/$\mbox{cm}^3$ and, in order to provide an upper bound to the generated entanglement that would be attainable using other materials, we shall use this density in our numerical simulations.
Accordingly, $\eta=1.36 \times 10^{-6}/\omega^2$, where $\omega$ is in Hz.

\begin{figure}[h]
\centering
\includegraphics[scale=0.35]{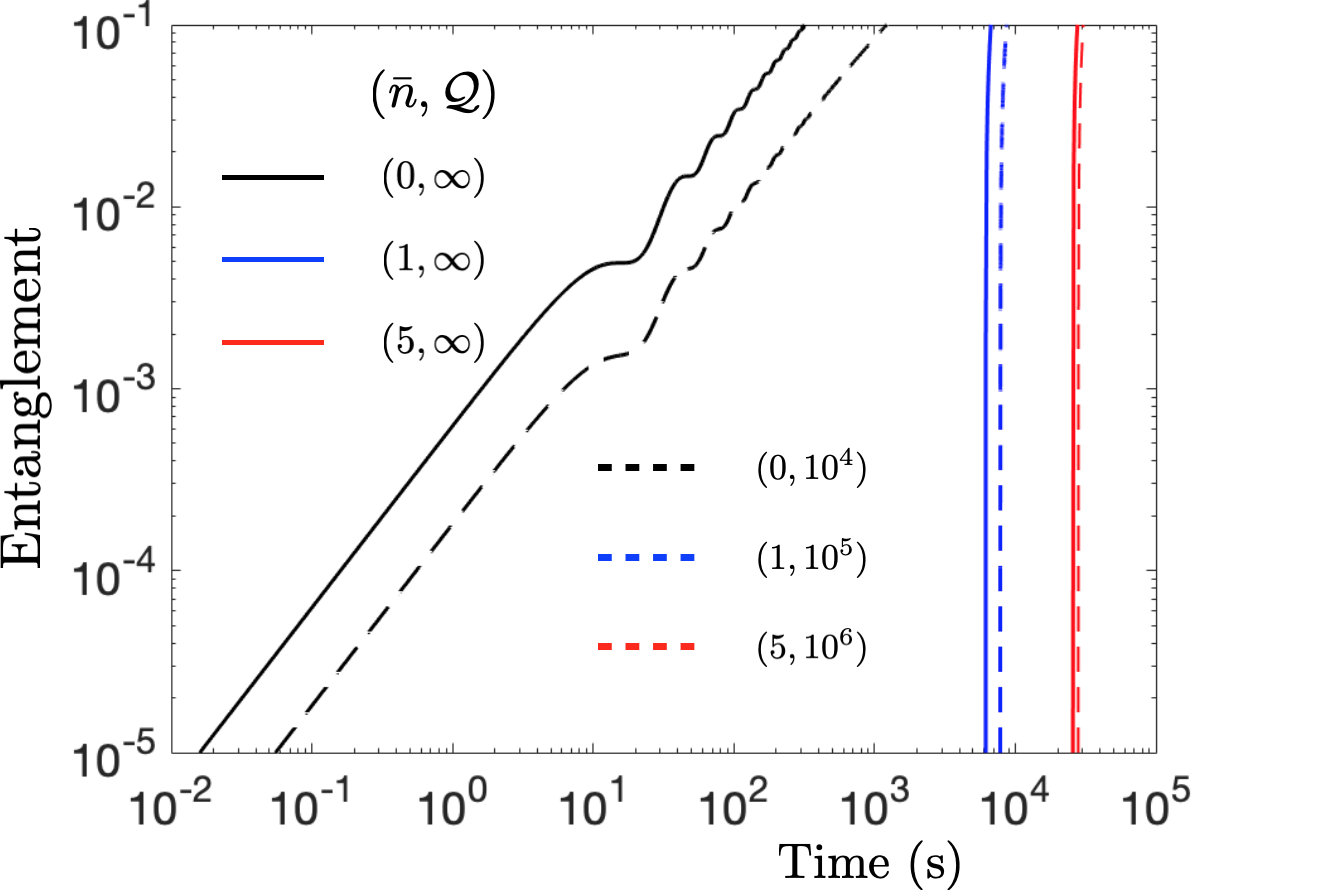}
\caption{Entanglement threshold and coherence time.
Different curves correspond to different pairs of parameters $(\bar n,\mathcal{Q})$, where $\bar n$ is the mean phonon number and $\mathcal{Q}$ the mechanical quality factor.
The frequency of both mirrors is taken to be $\omega=0.1$~Hz and the squeezing strength is $s_{A,B}=1.73$. The degree of entanglement is quantified using the logarithmic negativity.
}
\label{FIG_Edyndam}
\end{figure}

Fig.~\ref{FIG_Edyndam} shows exemplary entanglement dynamics for different values of the thermal phonon number $\bar n$ and mechanical quality factor $\mathcal{Q}$.
The frequency has been fixed to $\omega=0.1$~Hz (cf. Discussion section).
As expected, higher damping (lower $\mathcal{Q}$) results in the decay of entanglement, and the higher the temperature of the mirror (higher $\bar n$) the higher the mechanical quality factor needed to maintain entanglement. 
The setup allows for high entanglement, even with low coupling strength $\eta \sim 10^{-4}$.
However, this comes at the expense of the time for which the dynamics of the oscillators should be kept coherent.
It is also evident that cooling down the masses close to their ground state, $\bar n\approx 0$, is crucial for the reduction of the coherence time.\footnote{The oscillations of entanglement for unsqueezed initial state are still present in this dynamics, showing repeating pattern with period of $\pi/[(1-\eta)\omega ] \approx 31$~s.}

\vspace{0cm}
\noindent{\bf C. Calculations of entanglement: Released masses}

\noindent As seen, the experimental parameters required for detectable gravitational entanglement of masses in harmonic traps are demanding.
We therefore study one more feasible system, where the traps are switched off after cooling the masses.
Similar to the treatment of two oscillators, one starts with the total Hamiltonian for free masses and truncated gravitational term, and obtains the following equations of motion:
\begin{equation}\label{EQ_langevinsfm}
\begin{aligned}
\dot { X}_j &=\omega \:  P_j\qquad (j=A,B),\\ 
\dot { P}_A&=\omega \eta X_A-\omega\eta\, X_B+ \nu,\\
\dot { P}_B&=\omega \eta X_B-\omega\eta\, X_A-\nu.\\
\end{aligned}
\end{equation}
Note that $\omega$ here just sets the conversion between $x_j,p_j$ and their dimensionless counterparts $X_j,P_j$. 
In what follows, we will consider starting the dynamics with thermal state for each mass.
For example, the ground state is a Gaussian state with width $\Delta x(0)=\sqrt{\hbar/2m\omega}$.
This way, one can think of $\omega$ as a parameter characterising the initial spread of the wave function.

\begin{figure}[h]
\centering
\includegraphics[scale=0.35]{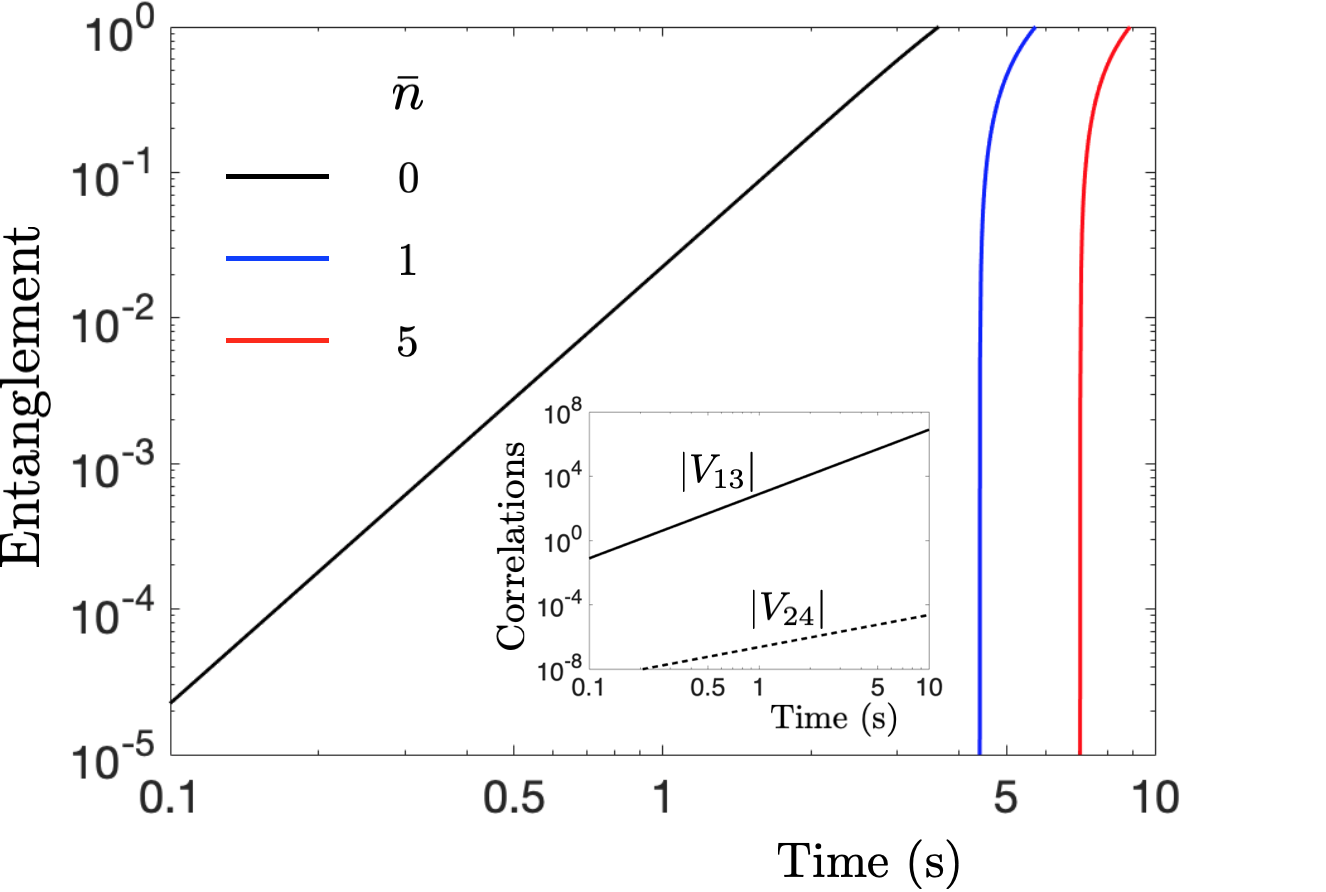}
\caption{Entanglement dynamics between two released Osmium spheres. 
Each mass has $m=100$~$\mu$g and is initially prepared in a Gaussian thermal state from a harmonic trap with $\omega=100$~kHz. The two masses are separated by $L=3R\approx 0.3$~mm. With these parameters gravity dominates Casimir interactions and observable entanglement is generated in seconds, which is shorter than the coherence times. 
In particular, entanglement in the order $0.01$ is achieved within $0.8$~s with initial ground state, and in $4.5$ ($7.5$)~s when starting with thermal states of $\bar n = 1$ ($5$). 
}
\label{FIG_fsdynamics}
\end{figure}

One can obtain the covariance matrix $V(t)$ from Eqs.~(\ref{EQ_langevinsfm}) and consequently derive the entanglement dynamics using the approach discussed in Methods section. 
After imposing the limits $\eta \ll 1$ and $\sqrt{\eta}\: \omega t \ll 1$, which apply in typical experimental situations, 
one obtains an analytical expression for the entanglement dynamics as follows:
\begin{eqnarray}\label{EQ_fment}
E_{\text {th}}(t)&=&\max \big \{0,E_{\text {gnd}}(t)-\log_2(2\bar n+1)\big \}, \\
E_{\text {gnd}}(t)&=&-\log_2\Big(\sqrt{1+2\sigma(t)-2\sqrt{\sigma(t)^2+\sigma(t)}}\Big),\nonumber
\end{eqnarray}
where $E_{\text {gnd}}(t)$ is the entanglement with initial ground state for each mass and $\sigma(t)=4G^2m^2\omega^2t^6/9L^6$.
Since entanglement is an increasing function of $\sigma(t)$, the latter is a figure of merit for entanglement gain relevant in the case of released masses. 
We present exemplary entanglement dynamics in Fig.~\ref{FIG_fsdynamics} for which entanglement $\sim10^{-2}$ is achieved within seconds. 
The parameters used here are $m=100$~$\mu$g, $\omega=100$~kHz, and $L=3R$.
We will show later that with these values gravity is the dominant interaction and coherence times are much longer than $1$~s.
Note that this setup does not require any squeezing.

These improvements over the scheme with trapped masses are the result of unlimited expansion of the wave functions. 
For example, for initial ground state, the evolution of the width of each sphere closely follows $\Delta x(t)\approx \sqrt{\hbar/2m\omega} \sqrt{1+\omega^2 t^2}$, which is an exact solution to a free non-interacting mass.
The effect of gravity is stronger attraction of parts of the spatial superposition that are closer, and hence generation of position and momentum correlations, leading to growing of quantum entanglement, see inset in Fig.~\ref{FIG_fsdynamics}.

In order to understand the effect of squeezing in this setup, let us suppose, for simplicity, the squeezing strengths $s_{A,B}=s$.
It is as if one initially prepared each mass in a Gaussian state with a new initial spread $\Delta x^{\prime}(0)=\Delta x(0)\exp(s)$. 
One can then calculate the entanglement dynamics using Eq. (\ref{EQ_fment}) with a new frequency $\omega^{\prime}=\omega \exp(-2s)$.
This means that anti-squeezing the initial position quadrature ($s>0$) would decrease entanglement gain, a situation opposite to the oscillators setup.
This is because a Gaussian state with smaller $\Delta x(0)$ spreads faster, such that during the majority of the evolution, the width is larger than that if one started with larger $\Delta x(0)$.
In principle, one obtains higher entanglement gain by squeezing the position quadrature ($s<0$).
However, this will result in higher final width, making it more susceptible to decoherence by environmental particles (see Discussion).

From numerical simulations, one confirms that, within $t=[0,10]$~s, the displacements of the two masses follow $x_A-x_B \ll L$. 
Furthermore, the trajectories coincide for both quantum treatment with truncated gravitational energy and classical treatment with full $H_{\text g}$ (see Supplementary Information).
This justifies the approximations used.

\vspace{0.5cm}
\noindent{\bf DISCUSSION}

\noindent We have shown that two nearby masses -- both trapped and released -- can become entangled via gravitational interaction.
Let us now discuss the conditions required to observe this entanglement in light of recent experimental achievements.

Logarithmic negativity in the order $10^{-2}$ has already been observed between mechanical motion and microwave cavity field~\cite{evalue1}. 
Extrapolating the same entanglement resolution to the case of two massive oscillators sets the required frequency to $\omega \sim 10^{-2}$~Hz, see Eq.~(\ref{EQ_OSC_MERIT}) \textcolor{black}{and its expression in terms of $\omega$}.
Interestingly, kilogram-scale mirrors of similar frequency ($\omega \sim 10^{-1}$~Hz) were recently cooled down near their quantum ground state~\cite{ligomirror}.
Furthermore, recent experiments on squeezed light have reported high squeezing strength \cite{sqpara1,sqpara2} (see also a review in this context \cite{sqpara3}), up to $15$~dB, which corresponds to $s \approx 1.73$.
Advances in the state transfer between light and optomechanical mirrors~\cite{optoreview} make this high squeezing promising also for mechanical systems.

For released masses, the experimental requirements are more relaxed. 
Their mass can be considerably smaller while the frequency for initial trapping considerably higher, which is close to common experimental parameters used for optomechanical system \cite{nanomirror,nm,optoreview}.
Note that higher frequency (lower $\Delta x(0)$) improves entanglement gain, unlike in the oscillators case where small $\omega$ is preferable.
However, one has to be cautious of decoherence mechanisms as a result of faster spreading rate of the wavefunctions. 
For future experiments, an improvement in the sensitivity of entanglement detection will also be beneficial. 

In light of their proximity, apart from gravitational interaction, the two masses can also interact via Casimir force.
It has been shown that the Casimir energy between two nearby spheres is given by a fraction of the ``proximity force approximation" $\mathcal{E}=-f_0 (\pi^3/1440)\hbar c R/(L - 2R + x_B - x_A)^2$, with the factor $0\le f_0\le1$ \cite{casimir1,casimir2}.
As typically $x_A - x_B \ll L - 2R$, we expand such expression to find a quadratic term in $x_A - x_B$ that can produce entanglement between the masses.
The strength of this term, however, is much weaker than the strength of the corresponding entangling term of gravitational origin:
for Osmium oscillators with mass $\sim 1$~kg separated by $L=2.1R$, the ratio between the Casimir and gravitational term is $r_{\text{cg}}\sim 10^{-12}$.
Similar calculations made for released masses of the same material with $m=100$~$\mu$g and $L=3R$, give $r_{\text{cg}}\sim 10^{-2}$.
It is thus legitimate to ignore Casimir interaction in both schemes.

Let us also discuss common decoherence mechanisms, i.e., due to interactions with thermal photons and air molecules \cite{decoherence}, see Supplementary Information for details.  
We take the average width of the wave function as an estimate for the superposition that is subjected to decoherence.
All the situations we consider follow the limit $\Delta x \ll \lambda$, i.e., the ``size'' of superposition is much smaller than the wavelength of the particles causing the decoherence. 
For oscillators made of Osmium, we use $m\sim 1$~kg and frequency $\omega \sim 0.1$~Hz.
Taking $L=2.1R$ and starting with ground state give $\Delta x \approx 8 \times 10^{-17}$~m.
From interactions with thermal photons at environmental temperature of $4$~K (liquid Helium), the coherence time for the oscillators is $\tau_{\text {ph}}\sim 5$~s.
The coherence time due to collisions with air molecules can be improved by evacuating the chamber with the oscillators -- for about $10^{12}$~molecules/$\mbox{m}^3$ (ultrahigh vacuum), the coherence time is $\tau_{\text {am}} \sim 5$~s.
One could also consider performing these experiments in space.
Taking the temperature as $2.7$~K (cosmic microwave background) and assuming $10^{7}$~molecules/$\mbox{m}^3$ (space pressure $\sim 10^{-15}$~Pa) we obtain $\tau_{\text {ph}}\sim 170$~s and $\tau_{\text {am}} \sim 10^6$~s.

By making similar calculations for released masses, with parameters considered in Fig.~3, one obtains $\tau_{\text {ph}}\sim 10^5$~s and $\tau_{\text {am}} \sim 10^{-4}$~s for the experiment on Earth with liquid Helium temperature and ultrahigh vacuum.
For the space experiment the coherence times improve to $\tau_{\text {ph}}\sim 10^7$~s and $\tau_{\text {am}} \sim 41$~s respectively.

Other schemes have been proposed for gravitationally induced entanglement~\cite{gravity1,gravity2}.
They are based on Newtonian interaction between spatially superposed microspheres with embedded spins.
\textcolor{black}{In those proposals, entanglement is reached faster and the small size of the experiment is the main advantage.
However, in order} to separate gravitational and Casimir contributions in that setup, each diamond sphere with mass $m=10^{-14}$~kg has to be superposed across 250~$\mu$m.
Decoherence due to scattering of molecules then becomes the main limiting factor.
The schemes we discussed here are complementary in a sense that vibrations of each oscillator are minute (no macroscopic superposition) but larger mass, $100$~$\mu$g, is required for observable entanglement.

We conclude with analysis of implications on the nature of the gravitational coupling that one can draw from such experiments and future research directions.
In laboratory, we deal with two nearby masses which are experimentally shown to become entangled.
These setups can be theoretically treated in different ways depending on the assumptions one makes about gravity.
In a ``conservative approach'' the two masses are coupled via Newtonian potential.
As seen from our calculations and those in Refs.~\cite{gravity1,gravity2} this indeed leads to gravitationally induced entanglement.
In this picture gravity is a direct interaction and hence it is difficult to draw conclusions about the form of quantumness of the gravitational field.
We note that even in this conservative approach such an experiment has considerable value as it would show the necessity of at least the quadratic term in the expansion of the Newtonian potential for generating entanglement.

The objection to the conservative approach is instantaneity of gravitational interaction: Newtonian potential directly couples masses independently of their separation.
\textcolor{black}{On the other hand, it has been shown that gravitational waves travel with finite speed~\cite{GravWaves}.}
For nearby masses this retardation is hardly measurable and Newtonian potential is \textcolor{black}{dominant and} expected to correctly describe the amount of generated entanglement.
A more consistent option in our opinion, motivated by quantum formalism and comparison with other fundamental interactions, is to treat gravitational field as a separate physical object.
In this picture the masses are not directly coupled, but each of them individually interacts with the field.
It has been argued within this mass-field-mass setting that entanglement gain between the masses implies non-classical features of the field~\cite{revealing,gravity1,gravity2}.

This discussion shows that it would be useful to provide methods for independent verification of the presence or absence of a physical object mediating the interaction.
We finish with a toy example of a condition capable of revealing that there was \emph{no} mediator.
To this end we consider two scenarios: (i) evolving a bipartite system described at time $t$ by a density matrix $\rho_{12}(t)$;
(ii) two objects interacting via a mediator $M$, i.e., with Hamiltonian $H_{1M} + H_{M2}$, described by a tripartite state $\rho_{1M2}(t)$.
We ask whether there exists bipartite quantum dynamics $\rho_{12}(t)$ that cannot be obtained by tracing out the mediator in scenario (ii).
Indeed, if $\rho_{12}(t)$ is a pure state at all times and entanglement increases, the dynamics could not have been mediated.
The purity assumption requires the mediator to be uncorrelated from $\rho_{12}(t)$, and uncorrelated mediator is not capable of entangling the principal system, composed of particles $1$ and $2$~\cite{revealing}.
It would be valuable to generalise this argument to mixed states measured at finite number of time instances.

\textcolor{black}{For example, in an experimental situation, the state of particles $1$ and $2$ might only be close to a pure state (with purity $1-\epsilon$, where $\epsilon$ is a small positive parameter) and therefore they could be weakly entangled to the mediator $M$ (with entanglement $\sim \mathcal{O}(\epsilon)$).
One would naturally expect that entanglement gain between particles $1$ and $2$ is bounded, e.g., as a function of $\epsilon$.
An observation of higher entanglement gain therefore excludes the possibility of mediated dynamics.}

\vspace{0.5cm}
\noindent{\bf METHODS}

\noindent{\bf A. Langevin equations and covariance matrix}

\noindent Let us first consider the setup with oscillators. 
One can rewrite the equations in (\ref{EQ_langevins}) as a single matrix equation $\dot u(t)=Ku(t)+l(t)$, with the vector $u(t)=(X_A(t), P_A(t), X_B(t), P_B(t))^T$ and a drift matrix
\begin{equation}\label{EQ_drift}
K=\left( \begin{array}{cccc} 
0&\omega&0&0\\ 
-\omega(1-\eta)&-\gamma&-\omega\eta&0\\
0&0&0&\omega\\
-\omega\eta&0&-\omega(1-\eta)&-\gamma
\end{array}\right).
\end{equation}
We split the last term in the matrix equation into two parts, representing the noise and constant term respectively, i.e., $l(t)=\upsilon(t)+\kappa$, where $\upsilon(t)=(0,\xi_A(t),0,\xi_B(t))^T$ and the constant $\kappa=\nu(0,1,0,-1)^T$ with $\nu={Gm^2}/\sqrt{\hbar m\omega L^4}$. 

The solution to the Langevin equations is given by 
\begin{eqnarray}\label{AEQ_Lsol}
u(t)&=&W_+(t)u(0)+W_+(t)\int_0^t dt^{\prime}  W_-(t^{\prime})l(t^{\prime}),
\end{eqnarray}
where $W_{\pm}(t)=\exp{(\pm Kt)}$. 
This allows one to calculate the expectation value of the $i$th quadrature $\langle u_i(t)\rangle$ numerically, which is given by the $i$th element of 
\begin{eqnarray}\label{AEQ_quad}
&&W_+(t)\langle u(0)\rangle+W_+(t)\int_0^t dt^{\prime}  W_-(t^{\prime})\kappa,
\end{eqnarray}
where we have used the fact that the noises have zero mean, i.e., $\langle \upsilon_i(t) \rangle=0$ and that $\langle \kappa \rangle=\mbox{tr}(\kappa \rho)=\kappa$.
From Eq.~(\ref{AEQ_Lsol}), one can also calculate other important quantities via the covariance matrix as shown below.

Covariance matrix of our system is defined as $V_{ij}(t)\equiv \langle \{ \Delta u_i(t),\Delta u_j(t)\}\rangle/2=\langle u_i(t)u_j(t)+u_j(t)u_i(t)\rangle/2-\langle u_i(t)\rangle \langle u_j(t)\rangle$ where we have used $\Delta u_i(t)=u_i(t)-\langle u_i(t)\rangle$. 
This means that $\kappa$ does not contribute to $\Delta u_i(t)$ (and hence the covariance matrix) since $\langle \kappa \rangle=\kappa$. 
We can then construct the covariance matrix at time $t$ from Eq. (\ref{AEQ_Lsol}) without considering $\kappa$ as follows:
\begin{eqnarray}
V_{ij}(t)&=&\langle u_i(t)u_j(t)+u_j(t)u_i(t)\rangle/2-\langle u_i(t)\rangle \langle u_j(t)\rangle \nonumber \\
\label{EQ_l3}
V(t)&=&W_+(t)V(0)W_+^T(t) \nonumber \\
&&+W_+(t)\int_0^t dt^{\prime} W_-(t^{\prime})DW_-^T(t^{\prime}) \:W_+^T(t) ,
\end{eqnarray}
where $D=\mbox{Diag}[0,\gamma(2\bar n+1),0,\gamma(2\bar n+1)]$ and we have assumed that the initial quadratures are not correlated with the noise quadratures such that the mean value of the cross terms are zero. 
A more explicit solution of the covariance matrix, after integration in Eq. (\ref{EQ_l3}), is given by 
\begin{eqnarray}\label{EQ_Ct}
KV(t)+V(t)K^T&=&-D+KW_+(t)V(0)W_+^T(t) \nonumber \\
&&+W_+(t)V(0)W_+^T(t)K^T \nonumber \\
&&+W_+(t)DW_+^T(t),
\end{eqnarray}
which is linear and can be solved numerically. 

Consider a special case, in which the damping term $\gamma$ is negligible, giving $D=\bm{0}$.
In this case, Eq. (\ref{EQ_Ct}) simplifies to 
\begin{equation}\label{EQ_nd}
V(t)=W_+(t)V(0)W_+^T(t).
\end{equation}
In this regime we have obtained analytical results in the main text.

For free masses, one can follow similar treatments as above, keeping in mind $\gamma=0$ and $\upsilon(t)=(0,0,0,0)^T$ such that the solution to quadrature dynamics and covariance matrix is given by Eq. (\ref{AEQ_quad}) and Eq. (\ref{EQ_nd}) respectively with a new drift matrix
\begin{equation}\label{EQ_driftfm}
K=\left( \begin{array}{cccc} 
0&\omega&0&0\\ 
\omega \eta&0&-\omega\eta&0\\
0&0&0&\omega\\
-\omega\eta&0&\omega \eta&0
\end{array}\right).
\end{equation}

\vspace{0.2cm}
\noindent{\bf B. Entanglement from covariance matrix}\label{APP_ENT}

\noindent The covariance matrix $V(t)$ describing our two-mode system can be written in a block form 
\begin{equation}
V(t)=\left( \begin{array}{cc} I_{A}&L\\ L^T&I_{B} \end{array}\right),
\end{equation}
where the component $I_{A}$ ($I_{B}$) is a $2\times 2$ matrix describing local mode correlation for $A$ ($B$) while $L$ is a $2\times 2$ matrix characterising the intermodal correlation. 
A two-mode covariance matrix has two symplectic eigenvalues $\{\nu_1,\nu_2\}$.
A physical system has $\nu_1,\nu_2\ge 1/2$ \cite{weedbrook2012gaussian}. 

For entangled modes, the covariance matrix will not be physical after partial transposition with respect to mode $B$ (this is equivalent to flipping the sign of the oscillator's momentum operator $P_B$ in $V(t)$). 
This unphysical $V(t)^{T_B}$ is shown by the minimum symplectic eigenvalue $\tilde \nu_{\min}<1/2$. 
The explicit expression is given by $\tilde \nu_{\min}=(\Sigma-\sqrt{\Sigma^2-4\:\mbox{det}V} )^{1/2}/\sqrt{2}$, 
where $\Sigma=\mbox{det}I_A+\mbox{det}I_B-2\:\mbox{det}L$.
Entanglement between mode $A$ and mode $B$ is then quantified by logarithmic negativity as follows $E=\max \big \{0,-\log_2{(2\tilde \nu_{\min})}\big \}$~\cite{negativity, adesso2004extremal}. 
Note that the separability condition, when $V(t)^{T_B}$ has $\tilde \nu_{\min}\ge1/2$, is sufficient and necessary for two-mode systems \cite{werner2001bound}. \\

\vspace{0.2cm}
\noindent{\bf ACKNOWLEDGEMENTS} 

\noindent We would like to thank Markus Aspelmeyer, Animesh Datta, Simon Gr\"{o}blacher, Su-Yong Lee, Nikolai Kiesel, Chiara Marletto, Vlatko Vedral, Harald Weinfurter, Marek \.Zukowski, Matteo Carlesso, Angelo Bassi, and Hendrik Ulbricht for stimulating discussions.
T.K. and T.P. thank Wies{\l}aw Laskowski for hospitality at the University of Gda{\'n}sk.
T.K. thanks Timothy C. H. Liew for hospitality at Nanyang Technological University.
This work is supported by the Singapore Ministry of Education Academic Research Fund Tier 2 Project No. MOE2015-T2-2-034 and Polish National Agency for Academic Exchange NAWA Project No. PPN/PPO/2018/1/00007/U/00001. 
G.Y.T. wishes to acknowledge the funding support for this project from Nanyang Technological University under the Undergraduate Research Experience on Campus (URECA) program.
M.P. is supported by the EU Collaborative Project TEQ (grant number 766900), the SFI-DfE Investigator Programme QuNaNet (grant number 15/IA/2864), the Leverhulme Trust Research Project Grant UltraQuTe (grant number RGP-2018-266), and the Royal Society Wolfson Research Fellowship RSWF/R3/183013.

\vspace{0.1cm}
\noindent{\bf Author contributions:} T.P. provided the initial project direction; T.K. carried out all the calculations and derivations under supervision of T.P. and M.P.; G.Y.T. assisted in confirming the derivations; T.K. wrote the paper with the help from T.P. and M.P. 

\vspace{0.1cm}
\noindent{\bf Competing interests:} The authors declare no competing interests.

\vspace{0.1cm}
\noindent {\bf Data availability:} All data needed to evaluate the conclusions in the paper are present in the paper and/or the Supplementary Information.

\clearpage

\newpage
\section*{SUPPLEMENTARY INFORMATION}

\noindent{\bf A. Details of entanglement dynamics: Oscillators}

\noindent In this section we show that entanglement gain is linked to the evolution of the position variance of each mass.
This is intuitive because bigger variance means stronger gravitational coupling for parts of the wave functions which are closer.
In order to illustrate this, we take, as an example, the oscillators setup with squeezed initial ground state for each mass.

\begin{figure}[h]
\includegraphics[scale=0.46]{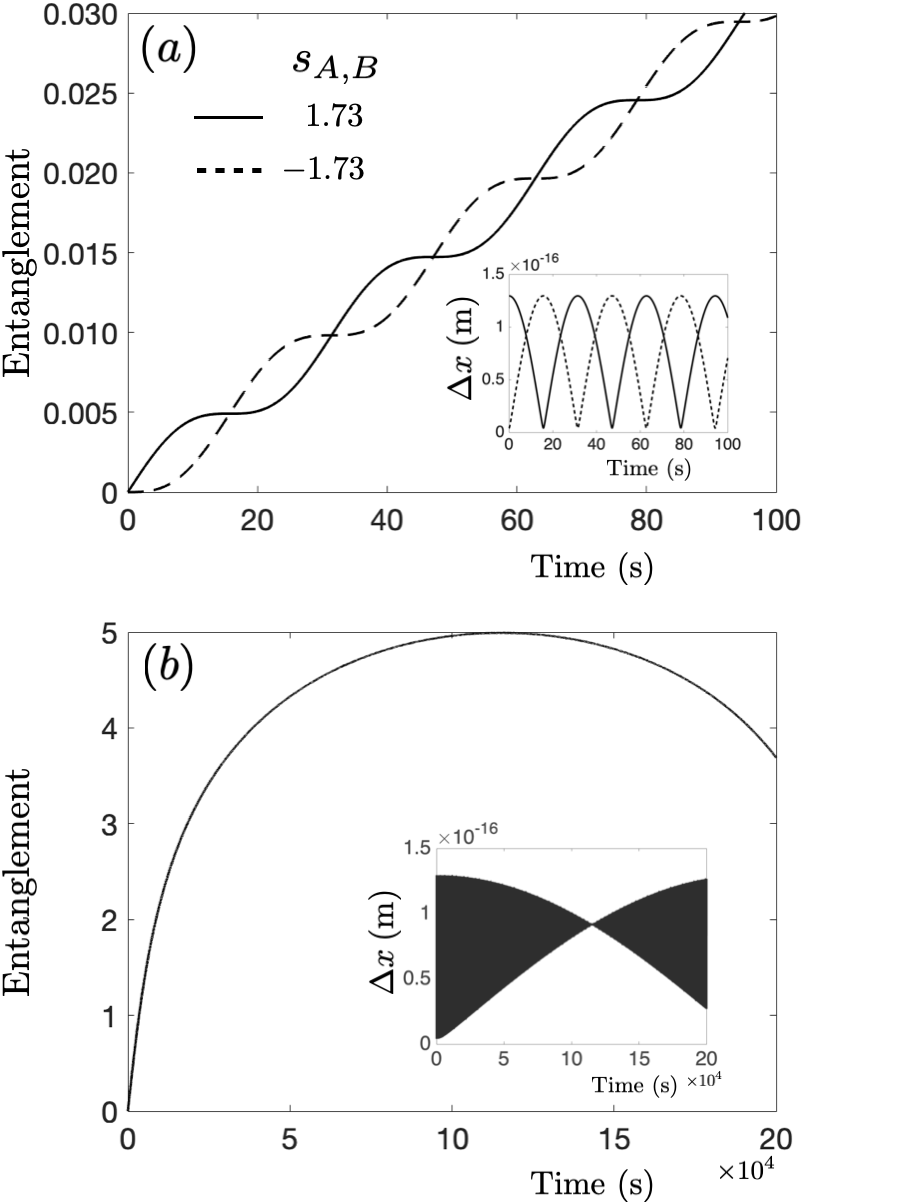}
\caption{Dynamics of entanglement and width of oscillators. 
The parameters used are $m=1$~kg with $\rho=22.59$~g/cm$^3$, $\omega=0.1$~Hz, and $L=2.1R$. (a)~The evolution is shown up to $100$~s for squeezing parameters $s_{A,B}=1.73$ and $-1.73$. (b)~Longer dynamics, showing the accumulation of entanglement. The dynamics are approximately the same for squeezed and anti-squeezed scenarios. 
}
\label{FIG_sqtogether}
\end{figure}

Fig. \ref{FIG_sqtogether}a shows the entanglement dynamics with both $s_{A,B}=1.73$ and $-1.73$, corresponding to initial states of the masses that are anti-squeezed and squeezed, respectively, in position quadrature.
During the evolution, the width of each mass oscillates as illustrated in the inset of Fig. \ref{FIG_sqtogether}a. 
It is clear that the oscillation of the width matches the oscillation of entanglement, for both squeezing parameters.
For longer time, i.e., $t\sim 10^4$~s, this leads to an accumulation of high entanglement as can be seen in Fig. \ref{FIG_sqtogether}b.
Note that the oscillation of the position variance is very rapid on this timescale and the envelope of these oscillations is the same for both positive and negative $s$, see the inset of Fig. \ref{FIG_sqtogether}b.
As a result, the entanglement dynamics is approximately equal for squeezed and anti-squeezed cases. 
This confirms our analytical result regarding the maximum entanglement gain being $|s_A + s_B|/\ln{2}$.


\vspace{0.5cm}
\noindent{\bf B. Quantum and classical trajectories: \\ Released masses}

\begin{figure}[!h]
\includegraphics[scale=0.25]{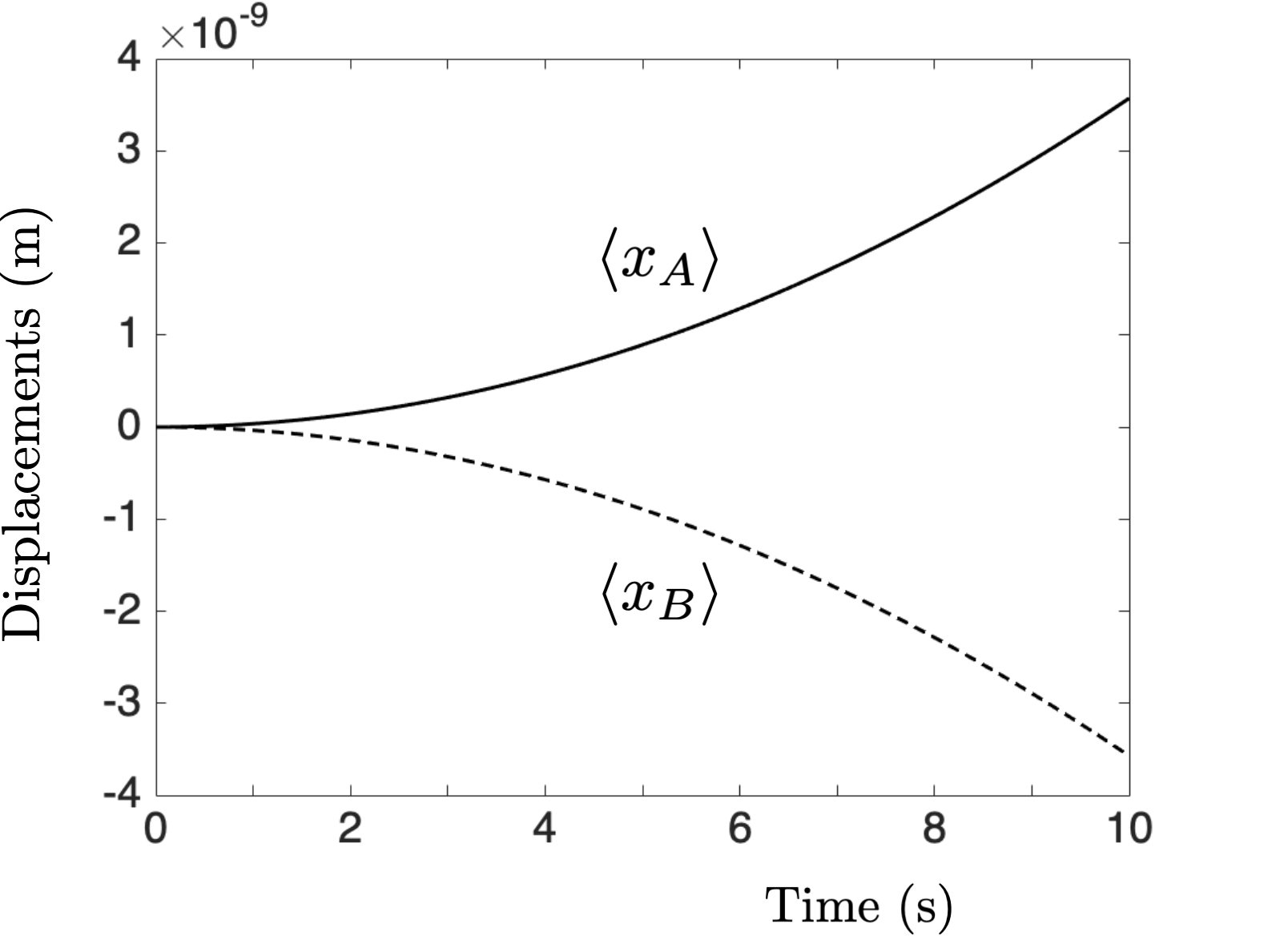}
\caption{Quantum trajectories of released masses coupled gravitationally. 
We assume Osmium spheres with $m=100$~$\mu$g and $L=3R$. Note that this dynamics is independent of the initial spreading parameter $\omega$ and the phonon number $\bar n$.
}
\label{FIG_traject}
\end{figure}

\noindent From the quadrature dynamics of Eq. (9) in the main text, one can calculate the expectation value of position for both masses. 
They are presented in Fig. \ref{FIG_traject}, where we have assumed initial conditions $\langle u(0)\rangle=(0,0,0,0)^T$.
On the other hand, without truncating the gravitational interaction, one can easily solve the classical dynamics and obtain the following equation for $x_t$:
\begin{equation}
t\sqrt{\frac{2Gm}{L}}=\sqrt{x_t(L-2x_t)}+\frac{L}{2\sqrt{2}}\left(\frac{\pi}{2}-\tan^{-1}(\theta(x_t))\right),
\end{equation}
where $\theta(x_t)=(L-4x_t)/\sqrt{8x_t(L-2x_t)}$ and $t$ is the time taken for the left mass to move a distance $x_t$. 
The trajectory of the right mass is simply $-x_t$.
One can confirm that the classical trajectories indeed coincide with the quantum ones.
This justifies the truncation of $H_{\text g}$ in our calculations. 
Note also that, within $10$~s, the displacement $(x_A-x_B)\sim 10^{-9}$~m, which is much smaller than the initial distance between the masses $L\approx 0.3$~mm.
This validates the use of the limit $x_A-x_B \ll L$.

\vspace{0cm}
\noindent {\bf C. Details of common decoherence}

\noindent Here we provide details about decoherence mechanisms due to interactions with thermal photons and air molecules \cite{decoherence}. 
In the regime where the superposition is much smaller than the wavelength of the scattering particles, i.e., $\Delta x \ll \lambda$, the coherence time due to interactions with thermal photons is given by $\tau_{\text {ph}}=1/\Lambda_{\text {ph}} (\Delta x)^2$ with
\begin{equation}
\Lambda_{\text {ph}}=10^{36} R^6\:T^9\: [1/\text m^2 \text s],
\end{equation}
where $R$ is the radius of the sphere and $T$ is the temperature of the environment. 
Note that all variables are in SI units.

The decoherence due to interactions with other scattering particles, e.g., air molecules, gives
\begin{equation}
\Lambda_{\text {am}}=\frac{8}{3\hbar^2}\frac{N}{V}\sqrt{2\pi m_{\text {air}}} \:R^2 (k_{\text B}T)^{3/2},
\end{equation}
where $N/V$ is the density of air molecules with mass $m_{\text {air}}$ and $k_{\text B}$ is Boltzmann constant.
We take $m_{\text {air}}\approx 0.5\times 10^{-25}$~kg.
For ultrahigh vacuum, pressure $\sim 10^{-10}$~Pa, the density is $\sim 10^{12}$~particles/m$^3$, while for a space experiment, pressure $\sim 10^{-15}$~Pa, the density can be as low as $\sim 10^7$~particles/m$^3$.

\vspace{0.5cm}
\noindent {\bf D. Different shapes of masses}\label{SUPP_diffcon}

\noindent First, let us note, from Eqs. (4) and (5), that the magnitude of the interaction rate between different modes in the case of identical spheres is given by $r_1= 2Gm/\omega L^3$. 
Assuming, e.g., the material is Osmium and $L=2.1R$, one gets $r_1=1.36\times 10^{-6}/\omega$.
Below, we will compare the interaction rate for different shapes of the objects to this reference.

\begin{figure}[!h]
\includegraphics[width=0.48\textwidth]{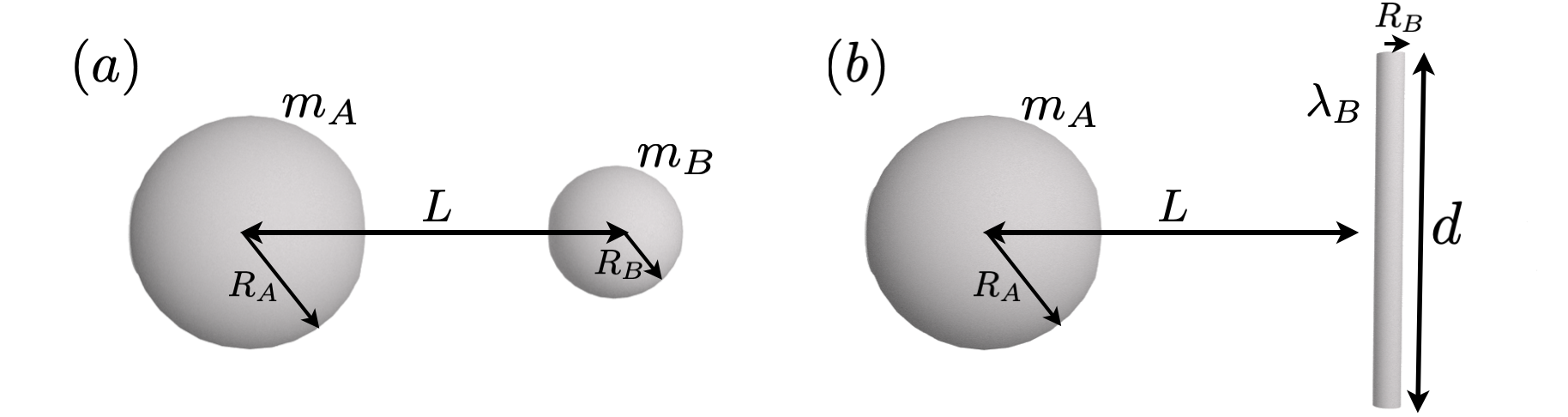}
\caption{Various shapes of the masses and notation used. 
}
\label{FIG_setupothers}
\end{figure}

Consider a particle of spherical shape, mass $m_A$, and frequency $\omega_A$ that interacts with a second sphere with mass $m_B$ and frequency $\omega_B$, see Fig. \ref{FIG_setupothers}a
(the same treatment applies to released masses with initial trapping frequencies $\omega_A$ and $\omega_B$).
Moreover, we assume both spheres are made of Osmium and take $R_B=\alpha R_A$, where $\alpha=[0,\infty )$.
After taking similar steps as those in the main text one obtains a new intermodal interaction rate $r_2= 1.36\times 10^{-6}f(\alpha)/\omega_A$, where $f(\alpha)=\alpha^{9/4}$ and we have used $L=2.1R_A$.
We have also assumed spring-like scaling for the frequency, i.e., $\omega_B=\omega_A\sqrt{m_A/m_B}$.
This result is intuitive as one expects stronger gravitational interaction by making the second sphere larger, i.e., larger $\alpha$.

More intriguing is the setting in Fig. \ref{FIG_setupothers}b, in which a rod with length $d$, mass $\lambda_B d$, and radius $R_B$ is interacting with a sphere of mass $m_A$ and radius $R_A$.
For simplicity, we assume both objects have a single mechanical frequency $\omega_A$ and $\omega_B$ respectively, and that the rod is thin such that its radius is much smaller than $L$. 
The gravitational interaction reads
\begin{equation}\label{EQ_hrodsphere}
H_{\text g}=-2Gm_A\lambda_B \ln{\left( \frac{d/2+\sqrt{(d/2)^2+(L^{\prime})^2}}{L^{\prime}}\right)},
\end{equation}
where $L^{\prime}=L-(x_A - x_B)$ with $x_A$ ($x_B$) being the displacement of mass $A$ ($B$). 
Note that this expression is the same as if we had a one-dimensional rod and a point mass.
By expanding Eq.~(\ref{EQ_hrodsphere}) in the limit $x_A - x_B \ll L$ and keeping only up to the quadratic term in displacements, one can show that the intermodal interaction rate is given by $r_3=2.18\times 10^{-7}f(\varsigma)/\omega_A$, where 
\begin{equation}
f(\varsigma)= (\varsigma)^{1/4}\left( 1-\frac{\varsigma^2((\varsigma^2-1)\sqrt{1+\varsigma^2}-1)}{(1+\sqrt{1+\varsigma^2})^2(1+\varsigma^2)^{3/2}}\right),
\end{equation}
and we define $\varsigma=2L/d$.
We have also taken $L=1.1R_A$, $R_B=0.1R_A$, and the same spring-like scaling as in the two-sphere configuration.
Although the strength of the gravitational energy (\ref{EQ_hrodsphere}) increases monotonically with $d$, it is not the case for $f(\varsigma)$, and hence $r_3$, 
which peaks at $\varsigma \approx 1.14$, i.e., $d\approx 1.75 L\approx 1.93R_A$.
The maximum $f_{\mbox{max}}=1.07$ gives maximum interaction rate $2.33\times 10^{-7}/\omega_A$.
For higher $d$ the rate $r_3$ decreases, implying that $r_3$ is always weaker than $r_1$.

We also note that both gravitational field and field gradient are necessary (but not sufficient) for producing entanglement.
This is clear from consideration of yet another configuration -- an infinite plane and a point mass separated by a distance $L$. 
One immediately observes that there is no field gradient here and that the gravitational energy of this configuration is proportional to $L-(x_A - x_B)$, which does not couple the masses and therefore does not contribute to entanglement.

\end{document}